\documentclass[]{spie}  %>>> use for US letter paper
\usepackage{amsmath,amsfonts,amssymb}
\usepackage{graphicx}
\usepackage{physics}
\usepackage[colorlinks=true, allcolors=blue]{hyperref}

\usepackage{algorithm}
\usepackage{algpseudocode}

\usepackage{amsmath}

\DeclareMathOperator*{\argmin}{arg\,min}
\usepackage{graphics}
\usepackage{epsfig}
\usepackage{multicol}

\title{An End-to-end Pipeline for 3D Slide-wise Multi-stain Renal Pathology Registration}

\author[a]{Peize Li*}
\author[a]{Ruining Deng*}
\author[a]{Yuankai Huo}
\affil[a]{Department of Computer Science, Vanderbilt University, Nashville, TN, 37235 USA}
\authorinfo{*Peize Li and Ruining Deng contribute equally to this paper\\  Corresponding author: Yuankai Huo: E-mail: yuankai.huo@vanderbilt.edu}

\begin{document} 
\maketitle

\begin{abstract}

 Tissue examination and quantification in a 3D context on serial section whole slide images (WSIs) were labor-intensive and time-consuming tasks. Our previous study proposed a novel registration-based method (Map3D) to automatically align WSIs to the same physical space, reducing the human efforts of screening serial sections from WSIs. However, the registration performance of our Map3D method was only evaluated on single-stain WSIs with large-scale kidney tissue samples. In this paper, we provide a Docker for an end-to-end 3D slide-wise registration pipeline on needle biopsy serial sections in a multi-stain paradigm. The contribution of this study is three-fold: (1) We release a containerized Docker for an end-to-end multi-stain WSI registration. (2) We prove that the Map3D pipeline is capable of sectional registration from multi-stain WSI. (3) We verify that the Map3D pipeline can also be applied to needle biopsy tissue samples. The source code and the Docker have been made publicly available at \href{https://github.com/hrlblab/Map3D} {{https://github.com/hrlblab/Map3D}}.

%  However, as the PIEC protocol is not designed for liver disease, typically only partial liver is included in the field-of-view (FOV). To tackle this challenge, our proposed HIPO pipeline employed the registration deformation to measure the volumetric variations of partial liver tissues. 

%  The measurement of volumetric variation of paired inspiratory-expiratory chest CT scans could be a promising method for liver disease diagnosis, such as liver cirrhosis. By means of liver segmentation and registration, we can accurately calculate the volume difference between inspiration and expiration, so that we can infer whether a specific liver is in good condition in terms of deformability. Through our study of 23 pairs of high resolution chest CT scans, we conclude that the absolute liver volumetric variation between expiration and inspiration is usually 5\%\--10\% ,which can be a benchmark applied in liver diagnosis.
\end{abstract}


\section{DESCRIPTION OF PURPOSE}
\label{sec:intro}  % \label{} allows reference to this section

\begin{figure}[t]
\begin{center}
\includegraphics[width=0.8\textwidth]{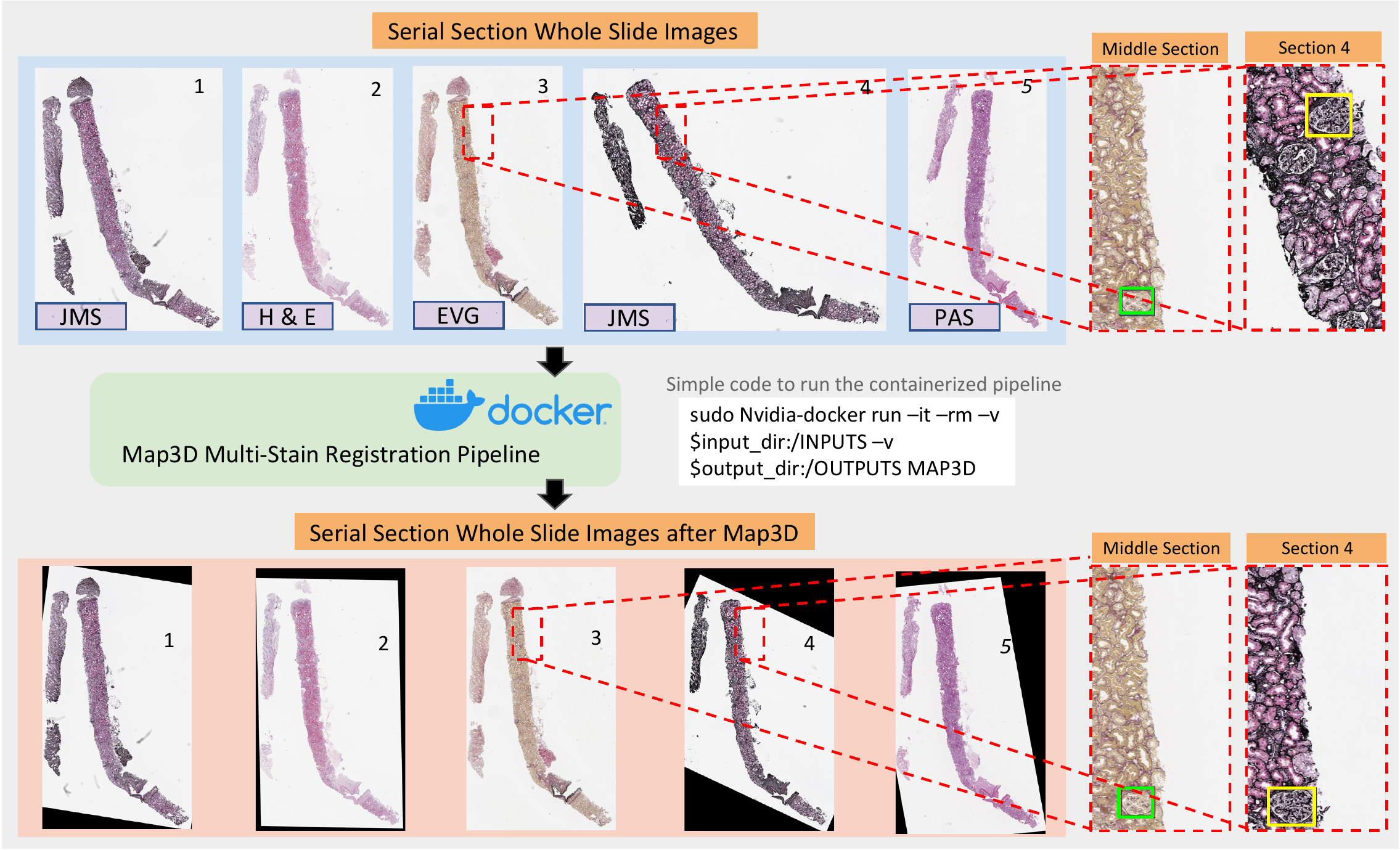}
\end{center}
\caption[example] 
{ \label{fig:stroy} 
This figure shows an overview of the Map3D registration pipeline. The input is a series of raw cross-section multi-stain WSIs, and the output is a series of images that have been transformed to the physical space of the middle section in a 3D context by transformation propagation.}
\end{figure} 
 
 %modern pathology analysis~\cite{aeffner2019introduction}
 
The tremendous advances in whole slide imaging and image processing techniques have led to a paradigm shift in modern pathology analysis. These advances are mostly attributed to deep-learning techniques, which empower computers to accomplish many labor-intensive and time-consuming manual tasks. However, the current quantitative assessments of primitive tissue structures across sectional WSIs still rely on the analysis of a single two-dimensional (2D) section and require manual assessments on WSI in 2D serial sectioning representation, leading to high labor costs and vulnerability. Therefore, large-scale 3D modeling in digital pathology is demanding for disease diagnosis. For example, the quantification of sclerotic areas within the capillary tuft has to be analyzed in 3D dimension to attain glomerular morphology for subtotal nephrectomy~\cite{remuzzi1995three}. Several disease-relative pathological changes, like Atubular, can only be confirmed when all WSI sections of a nephron are visually examined in 3D context\cite{chevalier2008generation}. Moreover, the 3D modeling in digital pathology can also make benefits to infuse additional phenotypic or functional information across the multi-staining biopsies when the adjacent biopsies are dyed by distinctive stains~\cite{roberts2012toward}.

To achieve large-scale 3D modeling in digital pathology, researchers have put much effort into developing models for whole series quantification. Robust biological image registration is critical to the performance of any 3D tissue quantification model. However, it is inevitable to have missing tissues and artifacts in WSI during section preparation, and these can lead to bad registration results~\cite{wright2020WSI}. Fortunately, our previously proposed deep-learning model Map3D ~\cite{deng2021map3d,deng2022dense} demonstrated a strong performance registering large-scale sectional WSIs in a 3D context, so it can further succeed in glomerular identification and association tasks that reduce the human efforts of screening serial sections from WSIs \footnote[2]{https://www.asn-online.org/education/kidneyweek/2020/program-abstract.aspx?controlId=3448300}. However, the previous Map3D registration model was only trained and tested on large-scale sectional WSIs, so we come up with the question that whether the registration capability can be kept for needle biopsy sections which contains less information and context. Moreover, the generalizability of Map3D registration to multi-stain sections also remained questionable.  

In this paper, we present an end-to-end 3D slide-wise registration pipeline on needle biopsy serial sections in a multi-stain paradigm: (1) We release a containerized Docker for an end-to-end multi-stain WSI registration. (2) We prove that the Map3D pipeline is capable of sectional registration from multi-stain WSI. (3) We verify that the Map3D pipeline can also be applied to needle biopsy tissue samples. The source code and the Docker have been made publicly available at \href{https://github.com/hrlblab/Map3D} {{https://github.com/hrlblab/Map3D}}.

\section{Method}
\begin{figure}
\begin{center}
\includegraphics[width=0.9\textwidth]{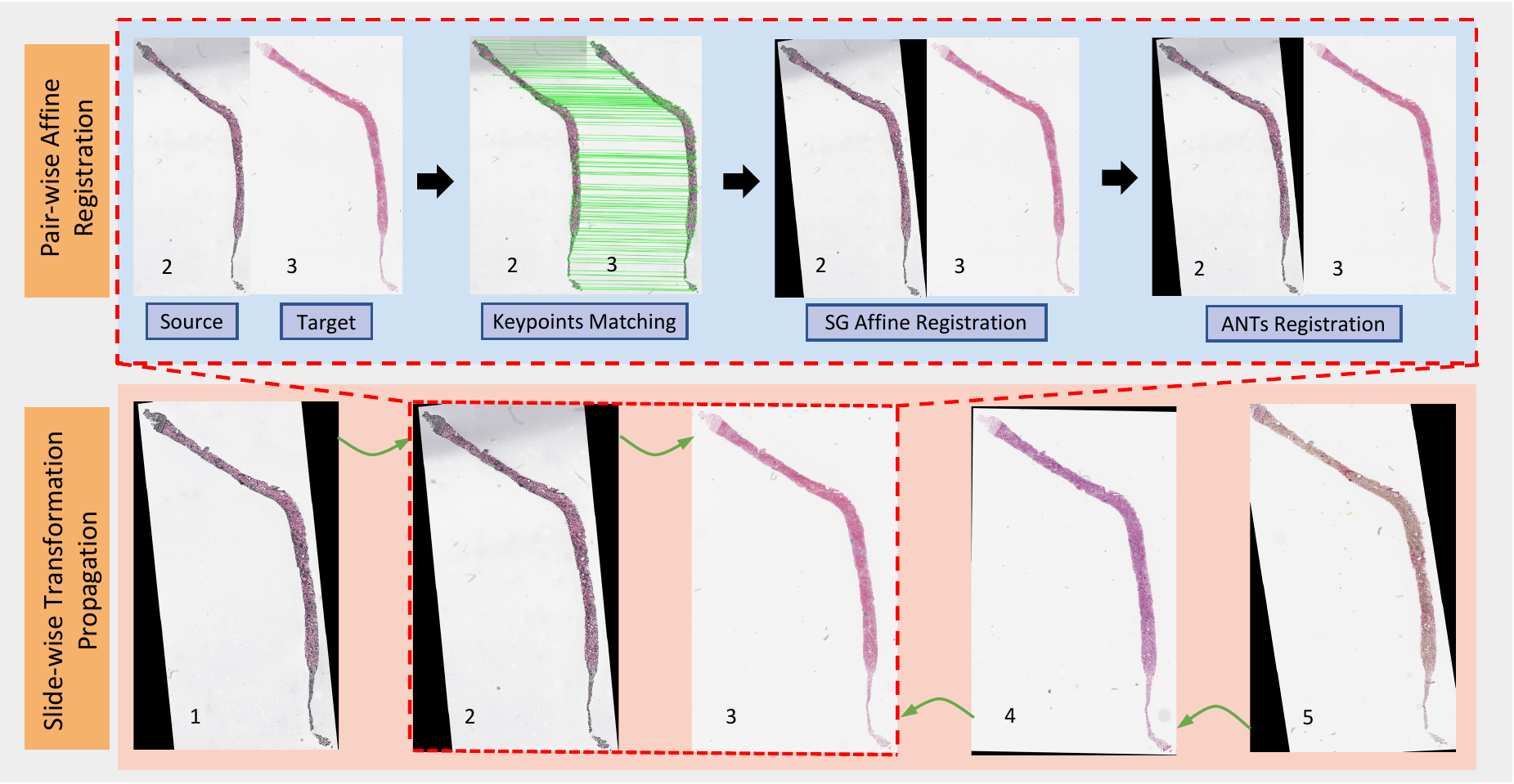}
\end{center}
\caption{This figure presents the whole pipeline of Map3D pair-wise affine registration, which is divided into 2 steps: (1)Two-stage pair-wise affine registration and (2) Slide-wise transformation propagation in a 3D context.}

\label{fig:method} 
\end{figure}

The overall end-to-end 3D registration pipeline is presented in Figure~\ref{fig:method}. The proposed pipeline consists of two sections: (1)Two-stage pair-wise affine registration and (2) Slide-wise transformation propagation in a 3D context.

\subsection{Two-stage pair-wise affine registration}
\label{sec:title}

Two-stage pair-wise registration is employed to find the pixel-to-pixel correspondence between different pathological images. To maintain the morphological characteristics of a tissue structure, we only deploy affine registration at each step rather than non-rigid registration. 

For the first step of the registration, recent Graph Neural Network (GNN) based registration SuperGlue (SG)~\cite{sarlin2020superglue} is employed to achieve keypoint-based global deformation on WSIs, since SG can filter out as well as match the similar key points on each pair of WSI with the reliable performance when encountering missing tissues and intensity changing~\cite{deng2021map3d}, significantly reducing the registration computing from pixel-wise registration of giga-pixel pathological images. The output of this step is a pair-wise affine matrix $M_{SG}(t)$ from Eq.(1).

\begin{equation}
M_{SG}(t) = \argmin \sum_{i=1}^N ||A(x_{i}^{t+1},M)-x_{i}^{t}||_{Aff_{SG}}
\label{SG}
\end{equation}

For the second step of the registration, advanced normalization tools (ANTs)~\cite{ants} is utilized due to the superior performance on high-resolution neuroimaging data. The output of this step is $M_{ANTs}(t)$ from Eq.(2).

\begin{equation}
M_{ANTs}(t) = \argmin A_{M_{SG}(t)} \sum_{i=1}^N ||A(x_{i}^{t+1},M)-x_{i}^{t}||_{Aff_{ANTs}}
\label{ANTs}
\end{equation}

We define $t$ as the $t$-th  section (frame) in the entire series with length $T-1$, where $i$ is the index of pixel $x_i$ in the image $I$, with $N$ pixels. The two-stage registration (SG + ANTs) affine matrix for each pair is presented in Eq.(3).

\begin{equation}
M(t) = (M_{SG}(t),M_{ANTs}(t))
\label{affine}
\end{equation}

In Eq.(1) and (2), $A$ indicates the affine registration. The affine matrix $M_{SG}(t)$ from SG is the initial parameter of ANTs. The $||.||_{Aff_SG}$ and $||.||_{Aff_ANTs}$ in Eq.(1) and (2) indicates the different similarity metrics for two affine registration, respectively. SG used Graph Neural Network layers to minimize the difference between each pair of keypoints, while ANTs registration was performed with default similarity metrics of mutual information (MI) and cross-correlation (CC), respectively. 

\subsection{Slide-wise transformation propagation in 3D context}
\label{sec:title}

After receiving pair-wise registration for all WSI pairs, we implement transformation propagation by doing sequence matrix multiplication to transfer sectional WSIs to the same physical space of the middle section $T_m$. The final transformation matrix in 3D context for $t$-th  section (frame) is shown in Eq(4).

\begin{equation}
M_{3D}(t) = 
\begin{cases}
(\prod_{i=t}^{T_m} M(i))^{-1},  \quad t < T_m \\
\prod_{i=T_m}^{t} M(i), \quad t > T_m 
\end{cases}
\end{equation}

where $(.)^{-1}$ is the inverse matrix of the multiplication.

\section{EXPERIMENTS}
\subsection{DATA}
\label{sec:title}

\begin{table*}[t]
\caption{Quantitative result for different baseline methods.}
\centering
\begin{tabular}{  l | c c c c  }
%\begin{tabular}{p{2cm}<p{0.25cm}<p{0.25cm}<p{0.25cm}<p{0.25cm}<p{0.25cm}<p{0.25cm}<p{0.25cm}<p{0.25cm}<p{0.25cm}<p{0.25cm}<p{0.25cm}<p{0.25cm}<p{0.25cm}<p{0.25cm}<p{0.5cm}<p{0.25cm}<p{0.25cm}}
%\begin{tabular}{p{3.0cm}p{1.0cm}p{1.0cm}p{1.0cm}p{1.0cm}p{1.0cm}p{1.0cm}}
%\begin{tabular}{p{2cm}<p{0.3cm}<p{0.3cm}<p{0.3cm}<p{0.3cm}<p{0.3cm}<p{0.3cm}<p{0.15cm}<p{0.15cm}<p{0.15cm}<p{0.15cm}<p{0.15cm}<p{0.15cm}<p{0.3cm}<p{0.3cm}<p{0.5cm}<p{0.5cm}<{0.5cm}}
%\begin{tabular}{p{2cm}<{\centering}p{0.4cm}<{\centering}p{0.4cm}<{\centering}p{0.4cm}<{\centering}p{0.4cm}<{\centering}p{0.4cm}<{\centering}p{0.4cm}<{\centering}p{0.15cm}<{\centering}p{0.2cm}<{\centering}p{0.2cm}<{\centering}p{0.2cm}<{\centering}p{0.2cm}<{\centering}p{0.2cm}<{\centering}p{0.3cm}<{\centering}p{0.3cm}<{\centering}|p{0.5cm}<{\centering}p{0.5cm}<{\centering}p{0.5cm}<{\centering}|}
 \hline
 Method& Distance (Mean)($\mu m $)& Distance (Median)($ \mu m $)& Box IoU (Mean)& Circle IoU (Mean) \\ 
 \hline
SG~\cite{sarlin2020superglue}& 65.27 & 40.87& 0.53 & 0.45\\ 
ANTs\cite{ants}& 372.48 & 23.50& 0.57 & 0.50\\
Map3D~\cite{deng2021map3d}& \textbf{36.41}& \textbf{19.71}& \textbf{0.65}& \textbf{0.58}\\ 
  
 \hline
\end{tabular}
\label{table1}
\end{table*}

\begin{figure}
\begin{center}
\includegraphics[width=0.9\textwidth]{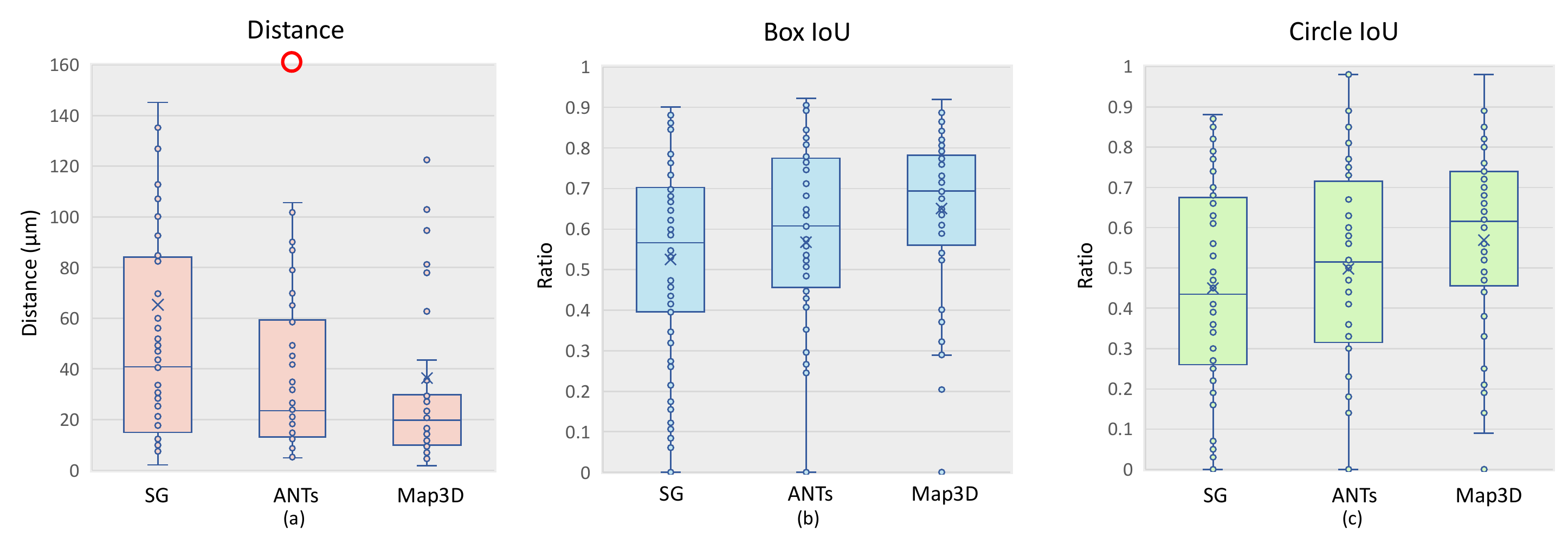}
\end{center}
\caption[example] 
{ \label{fig:plots} 
This figure shows the registration results from needle biopsy sections using SuperGlue, ANTs, and Map3D pipeline. (a) The average distance between a bounding box center on a non-middle section after registration and the corresponding bounding box center on the middle section. The red circle represents six extreme outliers from failure cases in ANTs registration. (b) The Intersection over Union between bounding boxes on the non-middle sections after registration and the corresponding bounding boxes on the middle sections. (c) The Intersection over Union between bounding circles on the non-middle sections after registration and the corresponding bounding circles on the middle sections. }
\end{figure} 

Image patches were extracted from 17 cases of renal transplant biopsies from Renal Pathology Teaching Collection by Virtual Pathology at the University of Leeds. This dataset contains serial sections of whole slide images in different kinds of stains, namely C4d, H\&E, CD45, Jones Methenamine Silver (JMS), Periodic Acid Shiff (PAS), Elastic Van Gieson (EVG), Polyoma Virus (PV), and Martius Scarlet Blue (MSB). All images were extracted from the 10 $\times$ magnification of a WSI with the original pixel resolution of 1 Micron.

For all sections, three glomeruli were selected as the region of interest and were manually annotated with rectangular bounding boxes on raw WSIs by ImageScope software. We saved the vertex coordinates for all the bounding boxes. In addition, we were confident that the selected glomeruli among all sections within the same case were derived from the same 3D object.

\subsection{Evaluation Metrics}
For the purpose of evaluating our results, we employed two types of metrics: the pair-wise distance from the middle section and the pair-wise Intersection over Union (IoU). The distance measures the physical distance between the center of a bounding box on a non-middle section after performing global registration and the center of a bounding box for the same glomerulus on the middle section. The IoU metric represents the level of overlap between these two bounding boxes.

Most of the glomeruli on the WSIs are in a ball-shape, which indicates that a significant part of a rectangular bounding box does not belong to our interested tissue. To reduce the effect of this error, we also included a special form of IoU metric that uses circles as the bounding region. A formula has been purposed to calculate the Intersection over Union of two circles given their centers and radii~\cite{circleNet}. To obtain these circles, we looked for the largest circle contained in a given bounding box. 

\section{Results}
Table~\ref{table1} presents the quantitative performance of our Map3D registration on multi-stain needle biopsy WSIs compared against two classical registration methods. Our previously proposed Map3D method achieved better performance than others in all evaluation metrics. According to Figure~\ref{fig:plots}, our two-stage registration pipeline keeps its superior power when generalize to multi-stain needle biopsy serial section WSIs with smaller registration errors and bigger overlaps. In addition, Map3D also demonstrates better robustness than the other two pipelines. As the table implies, although ANTs has a better overall performance than SuperGlue, it has several extreme outliers that leads to a larger average distance. In contrast, the result from Map3D is more consistent and reliable. All of these prove that our method also works with scarce information and context for needle biopsy tissue samples. 

\begin{figure}
\begin{center}
\includegraphics[width=0.95\textwidth]{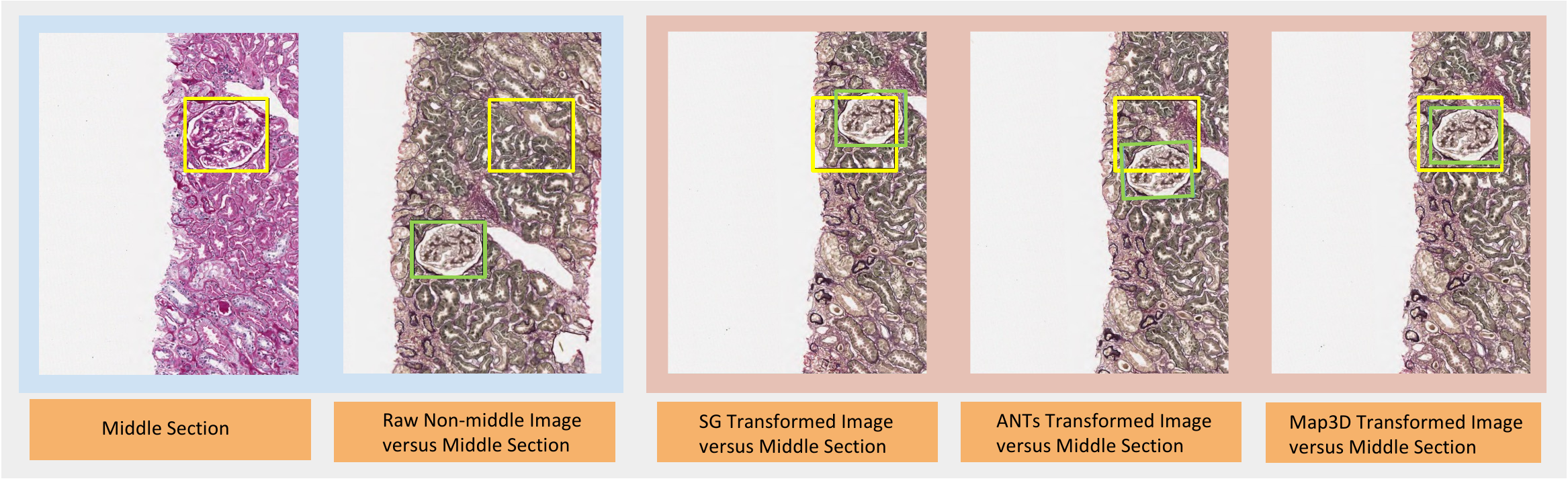}
\end{center}
\caption[example] 
{ \label{fig:results} 
This figure shows a comparison of an original section and its transformed sections from different registration models. The yellow bounding box represents the physical position of the glomerulus in the middle section, while the green bounding box represents the position of the glomerulus in a non-middle section before or after the registration. }
\end{figure}

An example of a glomerulus on a non-middle section getting registered can be found in Figure~\ref{fig:results}. The green box shows the position of a glomerulus on a non-middle section at the current stage, whereas the yellow box represents the physical position of the same glomerulus on the middle section. After Map3D Registration, the glomeulus was fitted into the yellow rectangle, while the other two registration method generated less ideal results.

%\begin{table*}
%\caption{Volume estimation performance.}
%\centering
%\begin{tabular}{p{3.0cm}p{0.5cm}p{0.5cm}p{0.5cm}p{0.5cm}p{0.5cm}p{0.5cm}p{0.5cm}p{0.25cm}p{0.25cm}p{0.25cm}p{0.25cm}p{0.25cm}p{0.25cm}p{0.25cm}p{0.5cm}p{0.5cm}p{0.5cm}}
%\begin{tabular}{cccc}
% \hline
% Method& Mean &Median&SD\\
% \hline
%Annotation&$1.99 \times 10^{6} $ & $1.94 \times 10^{6} $ & $7.23 \times 10^{5} $\\ 
%Our method&$2.71 \times 10^{6} $ & $2.76 \times 10^{6} $ & $7.38 \times 10^{5} $\\ 
%MPA~\cite{lane1992estimation}&$4.96 \times 10^{6} $ & $4.60 \times 10^{6} $ & $2.43 \times 10^{6} $\\ 
%2-profile~\cite{najafian2002estimating}&$6.30 \times 10^{6} $ & $3.63 \times 10^{6} $ & $6.43 \times 10^{6} $\\ %
% \hline
%\end{tabular}
%\label{table1}
%\end{table*}

%\begin{figure} 
%\begin{center}
%\includegraphics[height=8cm]{fig/VolumeResults.pdf}
%\end{center}
%\caption{This figure shows the relationships between annotations and volume estimations. The first row presents Spearman coefficients among different volume estimations. The second row presents the Bland-Altman plot among different volume estimations. The red circles are outliers, which significantly contribute to errors.}
%\label{fig:method} 
%\end{figure}
 
\section{NEW OR BREAKTHROUGH WORK TO BE PRESENTED}
\label{sec:breakthrough}
In this study, we release a Docker for an end-to-end 3D WSI registration pipeline. The entire process can now be run with one line of command. In addition, we also verified the robustness of our pipeline for registering needle biopsy tissue samples in the multi-stain paradigm. 

% \section{CONCLUSION}
% \label{sec:conclusion}
% According to our result, the Map3D performed a better registration task on 2D serial multi-stain needle biopsy whole slide images.

\section{ACKNOWLEDGMENTS}       
This work has not been submitted for publication or presentation elsewhere.

% References
\bibliography{main} % bibliography data in report.bib
\bibliographystyle{spiebib} % makes bibtex use spiebib.bst

\end{document}